\documentclass{article}
\usepackage{jamaica04}
\usepackage{epsfig}
\usepackage{graphicx}
\frompage{000} \topage{000}                                              
\def\rightmark{Spectra and correlations in p+p at RHIC}
\def\leftmark{T. Cs\"org\H{o} et al.}

\title{Analysis of identified particle yields and Bose-Einstein (HBT)
    correlations in p+p collisions at RHIC
}
\authors{
{T. Cs\"org\H{o}$^1$, M. Csan\'ad$^2$, B. L\"orstad$^3$ and A. Ster$^{1,4}$ %
}\\[2.812mm]
{\normalsize
\hspace*{-8pt}$^1$ MTA KFKI RMKI,
H-1525 Budapest 114, POBox 49, Hungary\\[0.2ex]
\hspace*{-8pt}$^2$ Dept. Atomic Physics, ELTE,
H-1117 Budapest, P\'azm\'any P. 1/a, Hungary \\
\hspace*{-8pt}$^3$ Dept. Physics, University of Lund,
S - 22362 Lund, Sweden\\[0.2ex]
\hspace*{-8pt}$^4$ MTA MFA,
H-1525 Budapest 114, POBox 49, Hungary\\[0.2ex]
}}

\abstract{
    Simultaneous Buda-Lund hydro model fits are presented to
    identified particle spectra and two-particle Bose-Einstein correlations
    as measured by the STAR collaboration in $\sqrt{s}=200$ GeV
    p+p collisions at RHIC. Preliminary results are compared to similar results
    in Au+Au collisions at RHIC, hadron+p reactions as well as
    Pb+Pb collisions at CERN SPS.
    }


\begin{document}

\maketitle
\setcounter{page}{1}

\section{Introduction}\label{intro}
      The Buda-Lund hydrodynamic model~\cite{3d,qm95}
    was shown to describe in a statistically acceptable
      manner the single particle spectra and the two-particle Bose-Einstein correlations
    in hadron+proton collisions at CERN SPS~\cite{na22}.
    The model, with different source parameters,
    also described well the identified single particle spectra and the two-particle
    Bose-Einstein correlations in  Pb+Pb collisions at CERN SPS~\cite{qm99}.
    Similarly, the Buda-Lund hydro model described well
    the single particle spectra and the two-particle Bose-Einstein
    correlation functions in Au+Au collisions at RHIC~\cite{Csanad:2004mm},
    both at $\sqrt{s_{NN}}=130$ and
    200 GeV colliding energies. The model was recently reviewed
    in ref.~\cite{cs-rev}
    and it was extended to describe elliptic flow in non-central heavy ion collisions
    at RHIC energies in refs.~\cite{Csanad:2003qa,Csanad:2004ci}.

    We have observed in these works,
    that the RHIC Au+Au data are well described
    by the model with a relatively high central temperature of
    214 $\pm$ 7 MeV in 0-10 \% central
    Au+Au collisions at $\sqrt{s_{NN}}=130$ GeV
    and 200 $\pm$ 9 MeV central temperature
    in the less central, 0-30 \% Au+Au collisions at  $\sqrt{s_{NN}}=200$ GeV.
    The same analysis suggests, that the surface (or with other words,
    the average) temperature of the hadronic final state is remarkably
    low, around $T_0 / 2$, nearly 100 MeV in these reactions.

    Lattice QCD calculations estimated the critical temperature of the
    hadron - quark gluon plasma phase transition near $\mu_B = 0$ to be
    $T_c = 172 \pm 3 $ MeV~\cite{Fodor:2001pe},
    which was recently calculated more precisely,
    using the physical quark masses, to be
    $T_c = 162 \pm  2 $ MeV~\cite{Fodor:2004nz}, even lower than thought before.
    Thus the Buda-Lund data analysis of central Au+Au spectra and correlation radii
    implies the existence of
    a superheated hadron gas in the very
    middle of the Au+Au collisions at RHIC, providing an indirect indication for the
    transition to deconfined matter~\cite{Csanad:2004mm}.
    However, at the lower CERN SPS energy, the central temperature
    was found, in  similar frameworks, to be within errors the smaller 139 $\pm$ 6 MeV,
    both in hadron+proton collisions~\cite{na22} and in Pb+Pb collisions~\cite{Csanad:2004mm},
    in line with the arguments of Landau that the freeze-out temperature
    of a hadron gas should be approximately the $\pi^0$ mass,
    the mass  of the lightest neutral quanta that can be excited in such a system.

    In this work we investigate two questions:

    a) Can a hydrodynamical model describe successfully the final state
    (identified particle spectra and Bose-Einstein correlations)
    in p+p collisions at RHIC?

    b) What are the best hydro model parameters and how they
    correspond to the values for hadron+p reactions at
    CERN SPS and Au+Au collisions at RHIC ?

    To answer these questions, we utilize as a tool the Buda-Lund hydrodynamical model.
    This model has been successfully applied to describe the double-differential
    invariant momentum distribution of charged particles in hadron+proton~\cite{na22}
    and Pb+Pb collisions~\cite{qm99} at CERN SPS fixed target experiments,
    as well in Au+Au collisions at  the RHIC collider energies
    of $\sqrt{s_{NN}} =  $ 130 GeV and 200 GeV~\cite{Csanad:2004mm}.
    Very few, if any, other models pass all of these tests (as far as we know).

\section{The Buda-Lund hydro model}\label{techno}

    The Buda-Lund hydro model was formulated in refs.~\cite{3d,qm95}, based
     on the following four principles:

    -- The particle emitting source consist of two, physically very different regions,
    the core and the halo. The core is a fireball like a star, the halo consists of the
    decay products of long-lived resonances, with relatively large length-scales,
    reminiscent to the solar wind surrounding the star. The large length-scales in the halo
    are assumed to be larger than the maximal length-scale resolvable by particle
    interferometry.

    -- The core is described by a locally thermalized, three dimensionally expanding
    particle emitting source. For central collisions, we utilize axial symmetry,
    for semi-central collisions, ellipsoidal symmetry to constrain the structure
    of the flow velocity, the fugacity and the local temperature distributions.

    -- We parameterize the hadronic final state in such a way, that it corresponds
    to know solutions of (relativistic or non-relativistic) hydrodynamical solutions
    in certain domains of the parameter space. In these limiting cases, even
    the time evolution of the local densities and the velocity field can be followed.

    -- We constrain the parameterization in such a way, that the observables can
    be calculated analytically from the model. This is useful, as simple scaling
    laws are observed experimentally in the transverse momentum spectra as well as
    in the Bose-Einstein (HBT) correlation functions. The Buda-Lund hydro model
    observes these scaling behaviors in certain domains of the parameter space.

\subsection{Hydro in soft hadron+hadron collisions}
\begin{figure}[htb]
\begin{center}
                 \includegraphics[width=2.4in]{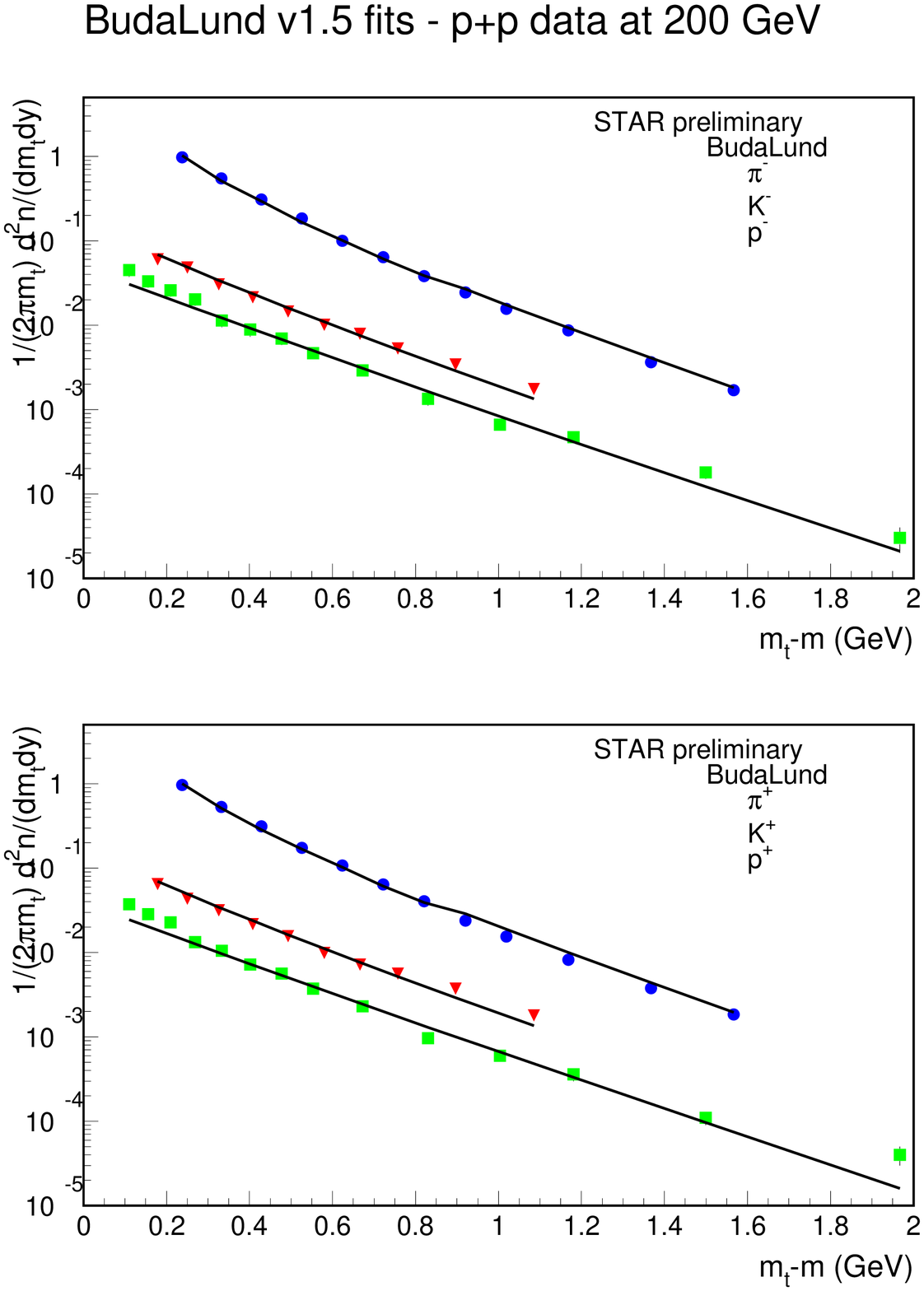}
                 \includegraphics[width=2.4in]{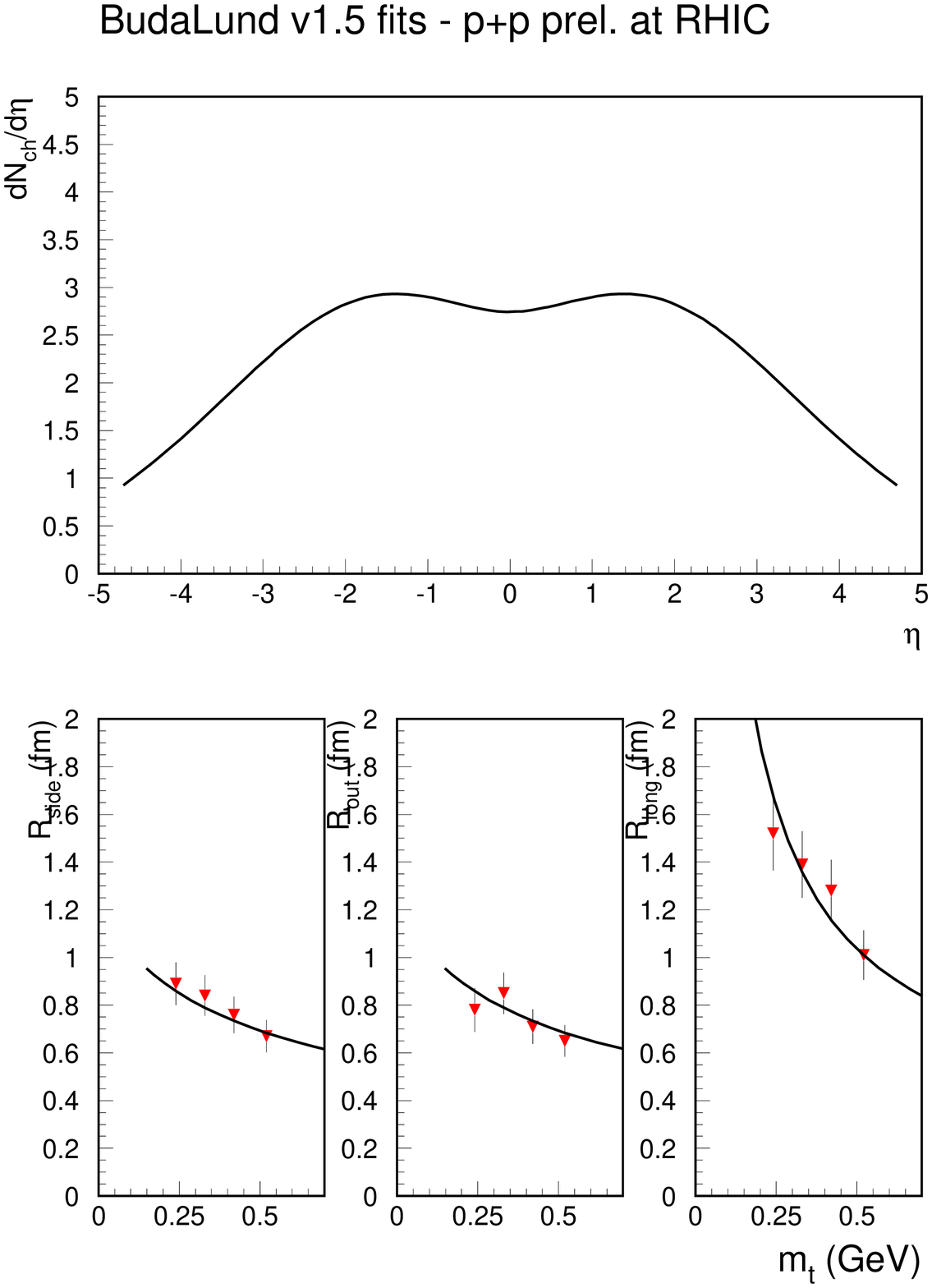}
\end{center}
\caption[]{
    The Buda-Lund hydro model is fitted here simultaneously to STAR
    single particle spectra\cite{STAR-pp-spectra-1,STAR-pp-spectra-2} and
    STAR preliminary two-pion correlation radius parameters~\cite{STAR-pp-radii}
    in p+p collisions at $\sqrt{s} =$ 200 GeV, with small amount of transverse flow,
    but with a large central temperature $T_0 \ge T_c$.
    The best fit parameters are summarized in Table I.
    The upper plot on the right side indicates the predicted pseudorapidity
    distribution of charged particles these collisions:
    BRAHMS and PHOBOS data can be used later, to restrict the source parameter $\Delta\eta$.
}
\label{fig1}
\end{figure}

    Recently, the STAR collaboration reported the identified particle spectra
    and the two-particle Bose-Einstein correlation radii in proton+proton collisions
    at the maximal RHIC colliding energy of $\sqrt{s_{NN}}=200$ GeV.
    At the Quark Matter 2004 conference, the transverse mass dependence of these
    radius parameters was presented as one of the new HBT puzzles at RHIC~\cite{Lisa-qm04}.
    However, the transverse mass dependence of the HBT radii in p+p collisions at RHIC is
    very similar to the transverse mass dependence of these radii in hadron+proton
    reactions at the lower CERN SPS energies~\cite{NA22-radii}, which were well described
    in terms of the Buda-Lund hydrodynamical model, together with the rapidity and transverse
    mass dependent double differential spectra of charged particles~\cite{na22}.
    In that case, the transverse mass dependence of these HBT radii was not generated by the
    transverse flow, but by the transverse temperature inhomogeneities of hadron+proton
    collisions.

    Perhaps it is surprising, that hydrodynamical descriptions work in the soft domain
    of high energy hadron+hadron collisions. However, Fermi and Landau initiated the
    statistical and hydrodynamical description of particle production for nucleon-nucleon
    as well as for nucleus-nucleus collisions. After accepting the hydrodynamical language,
    it is not surprising that the transverse temperature gradients play major role in
    hadron+hadron collisions - these systems have a rather small, typically 1 fm transverse
    radius, and the hot hadronic matter inside is surrounded with a vacuum that has zero
    temperature, so the temperature decreases on a small transverse radial scales,
    as compared to high energy heavy ion reactions, where the temperature decreases
    on a factor of 5-10 times larger scales.

\subsection{Buda-Lund fit results}

    In this section we show fit results to STAR final identified particle spectra
    ~\cite{STAR-pp-spectra-1,STAR-pp-spectra-2}
    and two-pion Bose-Einstein correlation radii~\cite{STAR-pp-radii}
    in  p+p  collisions at
     $\sqrt{s}=200$ GeV. First we investigate the question, is it possible
    to fit these data with the same central freeze-out temperature and transverse
    flow parameter, as were obtained~\cite{na22} in hadron+p collisions at CERN SPS?

    The answer is no, as also show graphically in the   presentation
    version of this work~\cite{csorgo-dnp04-talk}, however, this bad fit cannot
    be shown here due to space limitations. But the conclusion is clear:
    When fixing the freeze-out temperature in the center of the fireball
    to $T_0 = 140$ MeV and the transverse flow
    to $\langle u_t \rangle = 0.2 $, as in ref.~\cite{na22}, it is not possible
    to tune the other parameters of the Buda-Lund hydro model to describe the STAR p+p data
    at RHIC.

    However, the Buda-Lund hydro model can successfully describe all these
    observables in p+p collisions,
    when the central temperature and the transverse flow parameters
    are released, as shown in Fig. 1. From this comparison one learns, that the
    central temperature in p+p collisions at RHIC is significantly larger, than
    in h+p collisions at CERN SPS, although the transverse flow parameter is
    in both cases within errors compatible with zero.
    The fit parameters are summarized in Table 1.

\begin{table}[ht]
\begin{center}
\begin{tabular}{|l|rl|rl|rl|rl|rl|}
\hline
\hline
parameter
                & \multicolumn{2}{c|}{p+p \@ 200 GeV}
                & \multicolumn{2}{c|}{Au+Au \@ 200 GeV}
                & \multicolumn{2}{c|}{Au+Au \@ 130 GeV}
\\
                \hline
                \hline
$T_0$ [MeV]
                & \hspace{0.3cm} 289    &$\pm$ 8
                & \hspace{0.3cm} 200    &$\pm$ 9
                & \hspace{0.3cm} 214    &$\pm$ 7
                \\
$T_{\mbox{\rm e}}$ [MeV]
                & 90    &$\pm$ 42
                & 127    &$\pm$ 13
                & 102    &$\pm$ 11
                \\
$\mu_B$ [MeV]
                & 8     & $\pm$  76
                & 61    & $\pm$  40
                & 77    & $\pm$  38
                \\ \hline
$R_{G}$ [fm]
        & 1.2    &$\pm$ 0.3
        & 13.2   &$\pm$ 1.3
                & 28.0   &$\pm$ 5.5
                \\
$R_{s}$ [fm]
        & 1.1     &$\pm$ 0.2
        & 11.6    &$\pm$ 1.0
                & 8.6    &$\pm$ 0.4
                \\
$\langle u_t^\prime \rangle$
                & 0.04   &$\pm$  0.26
                & 1.5   &$\pm$  0.1
                & 1.0   &$\pm$  0.1
                \\ \hline
$\tau_0$ [fm/c]
        & 1.1    &$\pm$ 0.1
        & 5.7    &$\pm$ 0.2
                & 6.0    &$\pm$ 0.2
                \\
$\Delta\tau$ [fm/c]
        & 0.1    &$\pm$ 0.5
        & 1.9    &$\pm$ 0.5
                & 0.3    &$\pm$ 1.2
                \\
$\Delta\eta$
        & 3.0    &\mbox{\rm fixed}
        & 3.1    &$\pm$ 0.1
                & 2.4    &$\pm$ 0.1
                \\ \hline
$\chi^2/\mbox{\rm NDF}$
                & 89.7   &/ 69
                & 132    &/ 208
                & 158.2  &/ 180
                \\ \hline \hline
\end{tabular}
\caption
{Buda-Lund hydro model v1.5 source parameters, corresponding to Fig. 1.
The errors and the fit parameters are preliminary, as some of the fitted points
are not yet final and the statistical and systematic errors were
added in quadrature in these fits.
}
\label{tab:results}
\end{center}
\end{table}

\section{Conclusions}\label{concl}
    Let us quote Landau, who wrote~\cite{landau56} in 1956:
    ``Experiment shows that in collisions of very fast particles a
    large number of new particles are formed in multi-prong stars.
    The energy of the particles which produce such stars is of the  order of  $10^{12}$ eV
    or more.
    A characteristic feature is that such collisions occur not only  between
    a nucleon and a nucleus, but also between two nucleons."

    The Buda-Lund hydro model describes in a statistically acceptable manner
    the soft identified particle spectra and two-particle Bose-Einstein
    correlation data of STAR in p+p collisions at $\sqrt{s}=200$ GeV.
    The temperature distribution has a maximum which is significantly above
    the critical $T_c = 162 \pm 2 $ MeV value~\cite{Fodor:2004nz},
    but falls rapidly, within about 1.2 fm transverse
    distance, towards the vanishing temperature of the vacuum.

    Our results are in contrast to the recent statement by  Shuryak~\cite{Shuryakreview},
    who claimed that a hydrodynamical approach cannot describe the data
    in p+p collisions  at the RHIC energy domain.
    It seems to us that Landau's insight was more precise in this particular case.
    Even more surprising, the flow velocity profile, within the errors of
    reconstruction, corresponds to the Hwa-Bjorken famous 1+1 dimensional relativistic
    hydrodynamical solution~\cite{Hwa,Bjorken}. In the transversally more
    extended Au+Au collisions we find evidence for a fully developed,
    three dimensional Hubble flow~\cite{Csanad:2004mm}.
    For similar results,
     see~\cite{Cracow-spectra,Cracow-HBT,rel-cylsol,rel-ellsol,Retiere:2003kf}.

\section*{Acknowledgment(s)}
T. Cs. would like to thank the Organizers for creating a superb atmosphere
and organizing an inspiring and useful meeting.
We thank the STAR collaboration for making the
preliminary p+p $\rightarrow \pi + \pi + X $ HBT radii available.
  This work was supported by  following grants:
OTKA T034269, T038406, OTKA-MTA-NSF INT0089462,
NATO PST.CLG.980086, the exchange programmes
of Hungarian and Polish Academy of Sciences
and KBN-Hungarian Ministry of Education.

\end{document}